\documentclass[12pt]{revtex4}
\usepackage{amsmath,amssymb}
\usepackage{graphicx}
\usepackage{dcolumn}
\usepackage{bm}
\usepackage{revsymb}
\newcommand{\ext}{\operatorname{ext}}
\newcommand{\sagnac}{\operatorname{Sagnac}}
\newcommand{\In}{\operatorname{in}}
\newcommand{\out}{\operatorname{out}}

\begin{document}

\title{Quantum Theory in Accelerated Frames of Reference}

\author{Bahram Mashhoon}

\affiliation{Department of Physics and Astronomy\\
University of Missouri-Columbia\\
Columbia, Missouri 65211, USA}

\begin{abstract}\centerline{\bf{Abstract}}

The observational basis of quantum theory in accelerated systems is
studied. The extension of Lorentz invariance to accelerated systems
via the hypothesis of locality is discussed and the limitations of this
hypothesis are pointed out. The nonlocal theory of accelerated observers
is briefly described. Moreover, the main observational aspects of
Dirac's equation in noninertial frames of reference are presented.
The Galilean invariance of nonrelativistic quantum mechanics and the
mass superselection rule are examined in the light of the invariance
of physical laws under inhomogeneous Lorentz transformations.
\end{abstract}

\maketitle

\section{Introduction\label{sec:1}}

Soon after Dirac discovered the relativistic wave equation for a spin
$\frac{1}{2}$ particle~\cite{1}, the generally covariant Dirac equation
was introduced by Fock and Ivanenko~\cite{2} and was studied in great
detail by a number of authors~\cite{3}. Dirac's equation\begin{equation}
(i\hbar\gamma^{\alpha}\partial_{\alpha}-mc)\psi=0\label{eq:1}\end{equation}
transforms under a Lorentz transformation ${x'}^{\alpha}=L_{\;\;\beta}^{\alpha}\; x^{\beta}$
as\begin{equation}
\psi'(x')=S(L)\psi(x),\label{eq:2}\end{equation}
where $S(L)$ is connected with the spin of the particle and is given
by\begin{equation}
S^{-1}\gamma^{\alpha}S=L_{\;\;\beta}^{\alpha}\gamma^{\beta}.\label{eq:3}\end{equation}
The generally covariant Dirac equation can be written as \begin{equation}
(i\hbar\gamma^{\mu}\nabla_{\mu}-mc)\psi=0,\label{eq:4}\end{equation}
where $\nabla_{\mu}=\partial_{\mu}+\Gamma_{\mu}$ and $\Gamma_{\mu}$
is the spin connection. Let us consider a class of observers in spacetime
with an orthonormal tetrad frame $\lambda_{\;\;(\alpha)}^{\mu}$,
i.e.\begin{equation}
g_{\mu\nu}\lambda_{\;\;(\alpha)}^{\mu}\lambda_{\;\;(\beta)}^{\nu}=\eta_{(\alpha)(\beta)},\label{eq:5}\end{equation}
where $\eta_{(\alpha)(\beta)}$ is the Minkowski metric tensor.
Then in equation~\eqref{eq:4}, $\gamma^{\mu}$ is given by $\gamma^{\mu}=\lambda_{\;\;(\alpha)}^{\mu}\gamma^{(\alpha)}$
and\begin{equation}
\Gamma_{\mu}=-\frac{i}{4}\lambda_{\nu(\alpha)}[\lambda^{\nu}_{\;\; (\beta)}]_{;\mu}\; \sigma^{(\alpha)(\beta)},\label{eq:6}\end{equation}
where\begin{equation}
\sigma^{(\alpha)(\beta)}=\frac{i}{2}[\gamma^{(\alpha)},\gamma^{(\beta)}].\label{eq:7}\end{equation}
In this way, the generally covariant Dirac equation is minimally coupled
to inertia and gravitation.

The standard quantum measurement theory involves ideal inertial observers.
However, all actual observers are more or less accelerated. Indeed,
the whole observational basis of Lorentz invariance as well as quantum
mechanics rests upon measurements performed by accelerated observers.
It is therefore necessary to discuss how the measurements of noninertial
observers are connected with those of ideal inertial observers. This
paper is thus organized into two parts. In the first part, sections~\ref{sec:2}-\ref{sec:4},
we consider the basic physical assumptions that underlie the covariant
generalization of Dirac's equation. The second part, sections~\ref{sec:5}-\ref{sec:9},
are devoted to the physical consequences of this generalization for
noninertial frames of reference. In particular, the connection between
the relativistic theory and nonrelativistic quantum mechanics in accelerated
systems is examined in detail. Section~\ref{sec:10} contains a brief
discussion.

\section{Hypothesis of Locality\label{sec:2}}

The extension of Lorentz invariance to noninertial systems necessarily
involves an assumption regarding what accelerated observers actually
measure. What is assumed in the standard theory of relativity is the
\textit{hypothesis of locality,} which states that an accelerated
observer is pointwise equivalent to an otherwise identical momentarily
comoving inertial observer. It appears that Lorentz first introduced
such an assumption in his theory of electrons to ensure that an electron---conceived
as a small ball of charge---is always Lorentz contracted along its
direction of motion~\cite{4}. He clearly recognized that this is
simply an approximation based on the assumption that the time in which
the electron velocity changes is very long compared to the period
of the internal oscillations of the electron (see section 183 on page
216 of ~\cite{4}).

The hypothesis of locality was later adopted by Einstein in the course
of the development of the theory of relativity (see the footnote on
page 60 of ~\cite{5}). In retrospect, the locality assumption fits
perfectly together with Einstein's local principle of equivalence
to guarantee that every observer in a gravitational field is locally
(i.e. pointwise) inertial. That is, Einstein's heuristic principle
of equivalence, namely, the presumed local equivalence of an observer
in a gravitational field with an accelerated observer in Minkowski
spacetime, would lose its operational significance if one did not
know what accelerated observers measure. However, combined with the
hypothesis of locality, Einstein's principle of equivalence provides
a basis for a theory of gravitation that is consistent with (local)
Lorentz invariance.

Early in the development of the theory of relativity, the hypothesis
of locality was usually stated in terms of the direct acceleration
independence of the behavior of rods and clocks. The clock hypothesis,
for instance, states that ``standard'' clocks measure proper time.
Thus measuring devices that conform to the hypothesis of locality
are usually called ``standard''. It is clear that inertial effects
exist in any accelerated measuring device; however, in a \emph{standard}
device these effects are usually expected to integrate to a negligible
influence over the duration of each elementary measurement. Thus a
standard measuring device is locally inertial~\cite{6}.

Following the development of the general theory of relativity, the
hypothesis of locality was discussed by Weyl~\cite{7}. Specifically,
Weyl~\cite{7} noted that the locality hypothesis was an adiabaticity
assumption in analogy with slow processes in thermodynamics.

The hypothesis of locality originates from Newtonian mechanics: the
accelerated observer and the otherwise identical momentarily comoving
inertial observer have the same position and velocity; therefore,
they share the same \emph{state} and are thus pointwise identical
in classical mechanics. The evident validity of this assertion for
Newtonian point particles means that no new assumption is required
in the treatment of accelerated systems of reference in Newtonian
mechanics. It should also hold equally well in the classical \emph{relativistic}
mechanics of point particles, as originally recognized by Minkowski
(see page 80 of~\cite{8}). If all physical phenomena could be reduced
to \emph{pointlike coincidences} of particles and rays, then the hypothesis
of locality would be exactly valid.

The hypothesis of locality is not in general valid, however, in the
case of classical wave phenomena. Consider, for instance, the determination
of the frequency of an incident electromagnetic wave by a linearly
accelerated observer. Clearly, the frequency cannot be determined
instantaneously; in fact, the observer needs to measure a few oscillations
of the electromagnetic field before a reasonable determination of
the frequency becomes operationally possible. Let $\lambda$ be the
characteristic wavelength of the incident radiation and $\mathcal{L}$
be the acceleration length of the observer; then, the hypothesis of
locality is approximately valid for $\lambda\ll\mathcal{L}$. Here
$\mathcal{L}$ is a length scale that involves the speed of light
$c$ and certain scalars formed from the acceleration of the observer
such that the acceleration time $\mathcal{L}/c$ characterizes the
time in which the velocity of the observer varies appreciably. In
an Earth-based laboratory, for instance, the main translational and
rotational acceleration lengths would be $c^{2}/g_{\oplus}\approx1$ lt-yr
and $c/\Omega_{\oplus}\approx28$ AU, respectively. Thus in most experimental
situations $\lambda/\mathcal{L}$ is negligibly small and any possible
deviations from the locality hypothesis are therefore below the current
levels of detectability. Indeed, in the ray limit, $\lambda/\mathcal{L}\to0$,
the hypothesis of locality would be valid; therefore, $\lambda/\mathcal{L}$
is a measure of possible deviation from the locality postulate.

Consider a classical particle of mass $m$ and charge $q$ under the
influence of an external force $\mathbf{f}_{\ext}$. The accelerated
charge radiates electromagnetic radiation with a typical wavelength $\lambda
\sim\mathcal{L}$, where $\mathcal{L}$ is the acceleration length of the particle.We would
expect that a significant breakdown of the locality hypothesis occurs in this case,
since $\lambdabar /\mathcal{L}\sim 1$ in the interaction of the particle with the
electromagnetic field. The violation of the hypothesis of locality implies that the
state of the particle cannot be characterized by its position and velocity. This is
indeed the case, since the equation of motion of the radiating particle in the
nonrelativistic approximation is given by the Abraham-Lorentz equation
\begin{equation}\label{eq:8}m\frac{d\mathbf{v}}{dt}-\frac{2}{3}\frac{q^2}{c^3}\frac{d^2\mathbf{v}}{dt^2}+\dots
=\mathbf{f}_{\ext} \end{equation}
which implies that position and velocity are not sufficient to specify the state of the
radiating charged particle~\cite{9}.

To discuss quantum mechanics in an accelerated system of reference, it is therefore
useful to investigate the status of the hypothesis of locality vis-a-vis the basic
principles of quantum theory. The physical interpretation of wave functions is based on
the notion of wave-particle duality. On the other hand, the locality hypothesis is valid
for classical particles and is in general violated for classical waves. This
circumstance provides the motivation to develop a nonlocal theory of accelerated systems
that would go beyond the hypothesis of locality and would be consistent with
wave-particle duality. Such a theory has been developed~\cite{10} and can be employed,
in principle, to describe a nonlocal Dirac equation in accelerated systems of reference.
Some of the main aspects of the nonlocal theory are described in section~\ref{sec:4}.

\section{Acceleration Tensor\label{sec:3}}

It follows from the hypothesis of locality that an accelerated observer in Minkowski
spacetime carries an orthonormal tetrad  $\lambda^\mu_{\;\;(\alpha )}$, where
$\lambda^\mu_{\;\;(0)}=dx^\mu/d\tau $ is its four-velocity vector that is tangent to its
worldline and acts as its local temporal axis. Here $\tau$ is the proper time along the
worldline of the accelerated observer. To avoid unphysical situations, we assume
throughout that the observer is accelerated only for a finite period of time. The local
spatial frame of the observer is defined by the unit spacelike axes
$\lambda^\mu_{\;\;(i)}$, $i=1,2,3$. The tetrad frame is transported along the worldline
in accordance with
\begin{equation}\label{eq:9}\frac{d\lambda^\mu_{\;\;(\alpha)}}{d\tau
}=\Phi_\alpha^{\;\;\beta} \lambda^\mu_{\;\;(\beta)},\end{equation}
where
\begin{equation}\label{eq:10}\Phi_{\alpha\beta}=-\Phi_{\beta\alpha}\end{equation}
is the antisymmetric {\em acceleration tensor}. In close analogy with the Faraday
tensor, the acceleration tensor consists of ``electric" and ``magnetic" components. The
``electric" part is characterized by the translational acceleration of the observer such
that $\Phi_{0i}=a_i(\tau )$, where $a_i=A_\mu \lambda ^\mu _{\;\;(i)}$ and $A^\mu
=d\lambda^\mu_{\;\;(0)} /d\tau$ is the four-acceleration vector of the observer. The
``magnetic" part is characterized by the rotation of the local spatial frame with respect
to a locally nonrotating (i.e. Fermi-Walker transported) frame such that
$\Phi_{ij}=\epsilon_{ijk}\Omega^k$, where
$\bm{\Omega}(\tau )$ is the rotation frequency. The
elements of the acceleration tensor, and hence the spacetime scalars $\mathbf{a}(\tau )$
and $\bm{\Omega}(\tau )$, completely determine the local rate of variation of the state
of the observer. It proves useful to define the acceleration lengths $\mathcal{L}=c^2/a$
and $c/\Omega$, as well as the corresponding acceleration times $\mathcal{L}/c=c/a$ and
$1/\Omega$, to indicate respectively the spatial and temporal scales of variation of the
state of the observer. Let $\lambda$ be the intrinsic length scale of the phenomenon
under observation; then, we expect that the deviation from the hypothesis of
locality should be proportional to $\lambda /\mathcal{L}$.

It follows from a detailed analysis that if $D$ is the spatial dimension of a standard
measuring device, then $D\ll \mathcal{L}$~\cite{6}. Such devices are necessary for the
determination of the local frame of the accelerated observer. In fact, this circumstance
is analogous to the correspondence principle: while we are interested in the deviations
from the hypothesis of locality, such nonlocal effects are expected to be measured with
standard measuring devices.

\section{Nonlocality\label{sec:4}}

Imagine an accelerated observer in a background global Minkowski spacetime and let $\psi
(x)$ be a basic incident radiation field. The observer along its worldline passes
through a continuous infinity of hypothetical momentarily comoving inertial observers;
therefore, let $\hat{\psi} (\tau )$ be the field measured by the hypothetical inertial
observer at the event characterized by the proper time $\tau$. The local spacetime of
the hypothetical inertial observer is related to the background via a proper Poincar\'e
transformation $x'=Lx+s$; hence, $\psi'(x')=\Lambda (L)\psi(x)$, so that $\Lambda =1$
for a scalar field. We therefore assume that along the worldline $\hat{\psi}(\tau
)=\Lambda (\tau )\psi (\tau )$, where $\Lambda$ belongs to a matrix representation of
the Lorentz group.

Suppose that $\hat{\Psi}(\tau )$ is the field that is actually measured by the
accelerated observer. What is the connection between $\hat{\Psi}(\tau )$ and
$\hat{\psi}(\tau )$? The hypothesis of locality postulates the pointwise equivalence of
$\hat{\Psi}(\tau )$ and $\hat{\psi}(\tau )$, i.e. it requires that $\hat{\Psi}(\tau
)=\hat{\psi}(\tau )$. On the other hand, the most general linear relation between
$\hat{\Psi}(\tau )$ and $\hat{\psi}(\tau )$ consistent with causality is
\begin{equation}\label{eq:11} \hat{\Psi}(\tau )=\hat{\psi}(\tau
)+\int^\tau_{\tau_0}K(\tau ,\tau')\hat{\psi}(\tau ')d\tau ',\end{equation}
where $\tau_0$ is the initial instant of the observer's acceleration.
Equation~\eqref{eq:11} is manifestly Lorentz invariant, since it involves spacetime
scalars. The kernel $K(\tau ,\tau ')$ must be directly proportional to the observer's
acceleration, since $\hat{\Psi}=\hat{\psi}$ for an inertial observer. The
ansatz~\eqref{eq:11} differs from the hypothesis of locality by an integral over the past
worldline of the observer. In fact, this nonlocal part is expected to vanish for
$\lambdabar /\mathcal{L}\to 0$. The determination of a radiation field by an accelerated
observer involves a certain spacetime average according to equation~\eqref{eq:11} and
this circumstance is consistent with the viewpoint developed by Bohr and
Rosenfeld~\cite{11}.

Equation~\eqref{eq:11} has the form of a Volterra integral equation. According to
Volterra's theorem~\cite{12}, the relationship between $\hat{\Psi}$ and $\hat{\psi}$
(and hence $\psi$) is unique in the space of continuous functions. Volterra's theorem
has been extended to the Hilbert space of square-integrable functions by
Tricomi~\cite{13}.

To determine the kernel $K$, we postulate that a basic radiation field can never stand
completely still with respect to an accelerated observer. This physical requirement is a
generalization of a well-known consequence of Lorentz invariance to all observers. That
is, the invariance of Maxwell's equations under the Lorentz transformations implies that
electromagnetic radiation propagates with speed $c$ with respect to all inertial
observers. That this is the case for any basic radiation field is reflected in the
Doppler formula, $\omega '=\gamma (\omega-\mathbf{v}\cdot \mathbf{k})$, where $\omega
=c|\mathbf{k}|$. An inertial observer moving uniformly with speed $v$ that approaches $c$
measures a frequency $\omega '$ that approaches zero, but the wave will never stand
completely still $(\omega '\neq 0)$ since $v<c$; hence, $\omega '=0$ implies that
$\omega =0$. Generalizing this situation to arbitrary accelerated observers, we demand
that if $\hat{\Psi}$ turns out to be a constant, then $\psi$ must have been constant in the
first place. The Volterra-Tricomi uniqueness result then implies that for any true
radiation field $\psi$ in the inertial frame, the field $\hat{\Psi}$ measured by the
accelerated observer will vary in time. Writing equation~\eqref{eq:11} as
\begin{equation}\label{eq:12} \hat{\Psi}(\tau )=\Lambda(\tau )\psi (\tau
)+\int^\tau_{\tau_0}K(\tau ,\tau')\Lambda (\tau ')\psi (\tau ')d\tau',\end{equation}
we note that our basic postulate that a constant $\hat{\Psi}$ be associated with a
constant $\psi$ implies
\begin{equation}\label{eq:13} \Lambda (\tau _0)=\Lambda (\tau )+\int^\tau_{\tau_0}K(\tau
,\tau')\Lambda (\tau ')d\tau ',\end{equation}
where we have used the fact that $\hat{\Psi}(\tau_0)=\Lambda (\tau _0)\psi (\tau _0)$.
Given $\Lambda (\tau )$, equation~\eqref{eq:13} can be used to determine $K(\tau
,\tau')$; however, it turns out that $K(\tau ,\tau')$ cannot be uniquely specified in
this way. To go forward, it originally appeared most natural from the standpoint of
phenomenological nonlocal theories to postulate that $K(\tau ,\tau ')$ is only a
function of $\tau -\tau '$~\cite{10}; however, detailed investigations later revealed
that such a convolution kernel can lead to divergences in the case of nonuniform
acceleration~\cite{14}. It turns out that the only physically acceptable solution of
equation~\eqref{eq:13} is of the form~\cite{15,16}
\begin{equation}\label{eq:14} K(\tau ,\tau ')=k(\tau ')=-\frac{d\Lambda (\tau ')}{d\tau
'} \Lambda^{-1}(\tau ').\end{equation}
In the case of uniform acceleration, equation~\eqref{eq:14} and the convolution kernel
both lead to the same constant kernel. The kernel~\eqref{eq:14} is directly proportional
to the acceleration of the observer and is a simple solution of equation~\eqref{eq:13},
as can be verified by direct substitution. Moreover, if the acceleration of the observer
is turned off at $\tau_f$, then the unique kernel~\eqref{eq:14} vanishes for $\tau
>\tau_f$. Thus for $\tau >\tau_f$, the nonlocal contribution to the field in
equation~\eqref{eq:11} is simply a constant memory of the past acceleration of the
observer that is in principle measurable. This constant memory is simply canceled in a
measuring device whenever the device is reset.

For a scalar field $\Lambda =1$ and hence the kernel~\eqref{eq:14} vanishes. As will be
demonstrated in section~\ref{sec:8}, it follows from the locality of such a field that
for scalar radiation of frequency $\omega$, an observer rotating uniformly with
frequency $\Omega$ will measure $\omega '=\gamma (\omega -M\Omega)$, where $M=0,\pm 1,
\pm 2,\dots$. Thus $\omega '=0$ for $\omega =M\Omega$ and our basic physical postulate
is violated: the scalar radiation stands completely still for all observers rotating
uniformly about the same axis with frequency $\Omega$. It therefore follows from the
nonlocal theory of accelerated observers that a pure scalar (or pseudoscalar) radiation
field does not exist. Such fields can only be composites formed from other basic fields.
This consequence of the nonlocal theory is consistent with present observational data,
as they show no trace of a fundamental scalar (or pseudoscalar) field.

\subsection{Nonlocal Field Equations}

It follows from the Volterra equation~\eqref{eq:11} with kernel~\eqref{eq:14} that
\begin{equation}\label{eq:15} \hat{\psi} = \hat{\Psi} +\int ^\tau _{\tau_0} r(\tau ,\tau
') \hat{\Psi} (\tau ')d\tau ',\end{equation}
where $r(\tau ,\tau ')$ is the resolvent kernel. Imagine that a nonlocal field $\Psi$
exists in the background Minkowski spacetime such that an accelerated observer with a
tetrad frame $\lambda^\mu_{\;\;(\alpha )}$ measures
\begin{equation}\label{eq:16} \hat{\Psi}=\Lambda\Psi.\end{equation}
The relationship between $\Psi$ and $\psi$ can then be simply worked out using
\eqref{eq:15}, namely,
\begin{equation}\label{eq:17} \psi =\Psi +\int^\tau _{\tau_0} \tilde{r}(\tau ,\tau
')\Psi (\tau ')d\tau ',\end{equation}
where $\tilde{r}$ is related to the resolvent kernel by
\begin{equation}\label{eq:18} \tilde{r}(\tau ,\tau ')=\Lambda^{-1} (\tau )r(\tau ,\tau
')\Lambda (\tau ').\end{equation}

It is possible to extend equation~\eqref{eq:17} to a class of accelerated observers such
that $\psi(x)$ within a finite region of spacetime is related to a nonlocal field $\Psi
(x)$ by a suitable extension of equation~\eqref{eq:17}. The local field $\psi (x)$
satisfies certain partial differential equations; therefore, it follows from
\eqref{eq:17} that $\Psi$ would satisfy certain Lorentz-invariant nonlocal field
equations. In this way, the nonlocal Maxwell equations have been derived explicitly for
certain linearly accelerated systems~\cite{17}. It turns out that in general the field
equations remain nonlocal even after the cessation of accelerated motion.

\subsection{Nonlocal Electrodynamics}

To confront the nonlocal theory with observation, it is useful to derive the physical
consequences of nonlocal electrodynamics in systems that undergo translational and
rotational accelerations and compare the predictions of the theory with observational
data. It turns out that for accelerated systems the experimental data available at present 
do not have sufficient sensitivity to distinguish between the standard theory (based on
the locality hypothesis) and the nonlocal theory. In the case of linearly accelerated
systems, it may be possible to reach the desired level of sensitivity with the
acceleration of grains using high-intensity femtosecond lasers~\cite{18,19}. For a
uniformly rotating observer in circular motion, one can compare the predictions of
nonlocal electrodynamics with the nonrelativistic quantum mechanics of electrons in
circular atomic orbits or about uniform magnetic fields in the correspondence limit. If
the nonlocal theory corresponds to reality, its predictions should be closer to quantum 
mechanical results in the correspondence regime than those
of the standard local theory of accelerated systems. This turns out to be the case for
the simple cases that have been worked out in detail~\cite{20}.
Let us now return to the standard physical consequences of Dirac's equation in noninertial
systems of reference. In the following sections, emphasis will be placed on the main
inertial effects and their observational aspects in matter-wave interferometry.

\section{Inertial Properties of a Dirac Particle\label{sec:5}}

The physical consequences that follow from the Dirac equation in systems of reference
that undergo translational and rotational accelerations have been considered by a number
of authors~\cite{21}-\cite{24}. In particular, the work of Hehl and Ni~\cite{25} has
elucidated the general inertial properties of a Dirac particle. In their approach,
standard Foldy-Wouthuysen~\cite{26} transformations are employed to decouple the
positive and negative energy states such that the Hamiltonian for the Dirac particle may
be written as
\begin{equation}\label{eq:19} \mathcal{H} =\beta \left( mc^2+\frac{p^2}{2m}\right)
+\beta m\bm{a}\cdot \bm{x}-\bm{\Omega}\cdot (\bm{L}+\bm{S})\end{equation}
plus higher-order terms. Here $\beta m\bm{a}\cdot \bm{x}$ is an inertial term due to
the translational acceleration of the reference frame, while the inertial effects due to
the rotation of the reference frame are reflected in $-\bm{\Omega}\cdot (\bm{L}+\bm{S})$.

Before proceeding to a detailed discussion of these inertial terms in
sections~\ref{sec:6}-\ref{sec:9}, it is important to observe that Obukhov~\cite{27} has
recently introduced certain exact ``Foldy-Wouthuysen" (FW) transformations to decouple
the positive and negative energy states of the Dirac particle. Such a FW transformation
is defined up to a unitary transformation, which introduces a certain level of ambiguity
in the physical interpretation. That is, it is not clear from~\cite{27} what one could
predict to be the observable consequences of Dirac's theory  in noninertial systems and
gravitational fields. For instance, in Obukhov's exact FW transformation, an inertial
term of the form $-\frac{1}{2}\bm{S}\cdot \bm{a}$ appears in the Hamiltonian~\cite{27};
on
the other hand, it is possible to remove this term by a unitary transformation~\cite{27}.
The analog of this term in a gravitational context would be $\frac{1}{2}\bm{S}\cdot \bm{g}$.
Thus the energy difference between the states of a Dirac particle with spin polarized up
and down in a laboratory on the Earth would be $\frac{1}{2}\hbar g_\oplus \approx
10^{-23}$eV,
 which is a factor of five larger than what can be detected at present~\cite{28}. A
detailed examination of spin-acceleration coupling together with theoretical arguments
for its absence is contained in~\cite{29}.

The general question raised in~\cite{27} has been treated in~\cite{30}. It appears that
with a proper choice of the unitary transformation such that physical quantities would
correspond to simple operators, the standard FW transformations of Hehl and Ni~\cite{25}
can be recovered~\cite{30}. Nevertheless, a certain phase ambiguity can still exist in
the wave function corresponding to the fact that the unitary transformation may not be
unique. This phase problem exists even in the nonrelativistic treatment of quantum
mechanics in translationally accelerated systems as discussed in detail in
section~\ref{sec:9}.

\section{Rotation\label{sec:6}}

It is possible to provide a simple justification for the rotational inertial term in the
Hamiltonian~\eqref{eq:19}. Let us start with the classical nonrelativistic Lagrangian of
a particle $L=\frac{1}{2}mv^2-W$, where $W$ is a potential energy. Under a transformation to a
rotating frame of reference, $\bm{v}=\bm{v}'+\bm{\Omega}\times \bm{r}$, the Lagrangian
takes the form
\begin{equation}\label{eq:20}L'=\frac{1}{2}m(\bm{v}'+\bm{\Omega}\times\bm{r})^2-W,\end{equation}
where $W$ is assumed to be invariant under the transformation to the rotating frame. The
canonical momentum of the particle $\bm{p}'=\partial L'/\partial \bm{v}'=\bm{p}$ is an
invariant and we find that $H'=H-\bm{\Omega}\cdot \bm{L}$, where
$\bm{L}=\bm{r}\times\bm{p}$ is the invariant angular momentum of the particle. Let us
note that this result of Newtonian mechanics~\cite{31} has a simple relativistic
generalization: the rotating observer measures the energy of the particle to be
$E '=\gamma (E-\bm{v}\cdot \bm{p})$, where $\bm{v}=\bm{\Omega}\times \bm{r}$;
therefore, $E '=\gamma (E -\bm{\Omega}\cdot \bm{L})$.

This local approach may be simply extended to nonrelativistic quantum mechanics, where
the hypothesis of locality would imply that~\cite{32}
\begin{equation}\label{eq:21} \psi '(\bm{x}',t)=\psi (\bm{x},t),\end{equation}
since the rotating measuring devices are assumed to be locally inertial. Thus $\psi
'(\bm{x}',t)=R\psi(\bm{x}',t)$, where
\begin{equation}\label{eq:22} R=\hat{T}e^{\frac{i}{\hbar}\int^t_0 \bm{\Omega}(t')\cdot
\bm{J}dt'}.\end{equation}
Here $\hat{T}$ is the time-ordering operator and we have replaced $\bm{L}$ by
$\bm{J}=\bm{L}+\bm{S}$, since the {\em total} angular momentum is the generator of
rotations~\cite{32}. It follows that from the standpoint of rotating observers, $H\psi
=i\hbar \partial \psi /\partial t$ takes the form $H'\psi'=i\hbar \partial \psi'
/\partial t$, where
\begin{equation}\label{eq:23} H'=RHR^{-1}-\bm{\Omega }\cdot \bm{J}.\end{equation}
For the case of the single particle viewed by uniformly rotating observers, $H'$ can be
written as
\begin{equation}\label{eq:24} H'=\frac{1}{2m} (\bm{p}'-m\bm{\Omega}\times
\bm{r})^2-\frac{1}{2}m(\bm{\Omega}\times \bm{r})^2-\bm{\Omega}\cdot
\bm{S}+W,\end{equation}
where $-\frac{1}{2}m(\bm{\Omega}\times\bm{r})^2$ is the standard centrifugal potential
and $-\bm{\Omega}\cdot \bm{S}$ is the spin-rotation coupling term~\cite{32}. The
Hamiltonian~\eqref{eq:24} is analogous to that of a charged particle in a uniform
magnetic field; this situation is a reflection of the Larmor theorem. The corresponding
analog of the Aharonov-Bohm effect is the Sagnac effect for matter waves~\cite{33}. This
effect is discussed in the next section.

\section{Sagnac Effect\label{sec:7}}

The term $-\bm{\Omega}\cdot \bm{L}$ in the Hamiltonian~\eqref{eq:19} signifies the coupling
of the orbital angular momentum of the particle with the rotation of the reference frame
and is responsible for the Sagnac effect exhibited by the Dirac particle. The
corresponding Sagnac phase shift is given by
\begin{equation}\label{eq:25} \Delta \Phi_{\sagnac}=\frac{2m}{\hbar}
\int\bm{\Omega}\cdot d\bm{A},\end{equation}
where $\bm{A}$ is the area of the interferometer. Equation~\eqref{eq:25} can be
expressed as
\begin{equation}\label{eq:26} \Delta \Phi_{\sagnac} =\frac{2\omega}{c^2}\int
\bm{\Omega}\cdot d\bm{A},\end{equation}
where $mc^2\approx \hbar \omega$ and $\omega$ is the de Broglie frequency of the
particle. Equation~\eqref{eq:26} is equally valid for electromagnetic radiation of
frequency $\omega$.

For matter waves, the Sagnac effect was first experimentally measured for Cooper pairs
in a rotating superconducting Josephson-junction interferometer~\cite{34}. Using slow
neutrons, Werner et al.~\cite{35} measured the Sagnac effect with $\bm{\Omega}$ as the
rotation frequency of the Earth. The result was subsequently confirmed with a rotating
neutron interferometer in the laboratory~\cite{36}. Significant advances in atom
interferometry have led to the measurement of the Sagnac effect for neutral atoms as
well. This was first achieved by Riehle et al.~\cite{37} and has been subsequently
developed with a view towards achieving high sensitivity for atom interferometers as
inertial sensors~\cite{38}. In connection with charged particle interferometry, the
Sagnac effect has been observed for electrons by Hasselbach and Nicklaus~\cite{39}.

The Sagnac effect has significant and wide-ranging applications and has been reviewed
in~\cite{40}. 

\section{Spin-Rotation Coupling\label{sec:8}}

The transformation of the wave function to a uniformly rotating system of coordinates
involves $(t,r,\theta ,\phi )\to (t,r,\theta ,\phi +\Omega t)$ in spherical coordinates,
where $\Omega$ is the frequency of rotation about the $z$ axis. If the dependence of the
wave function on $\phi$ and $t$ is of the form $\exp (iM\phi -iEt/\hbar )$, then in the
rotating system the temporal dependence of the wave function is given by $\exp
[-i(E-\hbar M\Omega )t/\hbar ]$. The energy of the particle measured by an observer at
rest in the rotating frame is
\begin{equation}\label{eq:27} E'=\gamma (E -\hbar M\Omega ),\end{equation}
where $\gamma =t/\tau$ is the Lorentz factor due to time dilation. Here $\hbar M$ is the {\em
total} angular momentum of the particle along the axis of rotation; in fact, $M=0,
\pm 1 ,\pm 2,\dots$, for a scalar or a vector particle, while $M\mp \frac{1}{2}=0,\pm 1,
\pm 2,\dots$, for a Dirac particle.

In the JWKB approximation, equation~\eqref{eq:27} may be expressed as $E'=\gamma
(E-\bm{\Omega }\cdot \bm{J})$ and hence
\begin{equation}\label{eq:28} E'=\gamma (E-\bm{\Omega }\cdot \bm{L})-\gamma \bm{\Omega
}\cdot \bm{S}.\end{equation}
It follows that the energy measured by the observer is the result of an instantaneous
Lorentz transformation together with an additional term
\begin{equation}\label{eq:29} \delta H=-\gamma \bm{\Omega}\cdot \bm{S},\end{equation}
which is due to the coupling of the intrinsic spin of the particle with the frequncy of
rotation of the observer~\cite{32}. The dynamical origin of this term can be simply
understood on the basis of the following consideration: The intrinsic spin of a free
particle remains fixed with respect to the underlying global inertial frame; therefore, 
 from the standpoint of observers at rest in the rotating system, the spin precesses in
the opposite sense as the rotation of the observers. The Hamiltonian responsible for
this inertial motion is given by equation~\eqref{eq:29}. The relativistic nature of
spin-rotation coupling has been demonstrated by Ryder~\cite{41}. Let us illustrate these
ideas by a thought experiment involving the reception of electromagnetic radiation of
frequency $\omega$ by an observer that rotates uniformly with frequency $\Omega$. We
assume for the sake of simplicity that the plane circularly polarized radiation is
normally incident on the path of the observer, i.e. the wave propagates along the axis
of rotation. We are interested in the frequency of the wave $\omega '$ as measured by
the rotating observer. A simple application of the hypothesis of locality leads to the
conclusion that the measured frequency is related to $\omega$ by the transverse Doppler
effect, $\omega '_D=\gamma \omega$, since the instantaneous rest frame of the observer is
related to the background global inertial frame by a Lorentz transformation. On the
other hand, a different answer emerges when we focus attention on the measured
electromagnetic field rather than the propagation vector of the radiation,
\begin{equation}\label{eq:30} F_{(\alpha )(\beta )}(\tau
)=F_{\mu\nu}\lambda^\mu_{\;\;(\alpha )} \lambda^\nu_{\;\;(\beta )},\end{equation}
where $F_{\mu\nu}$ is the Faraday tensor of the incident radiation and
$\lambda^\mu_{\;\;(\alpha )}$ is the orthonormal tetrad of the rotating observer. The
nonlocal process of Fourier analysis of $F_{(\alpha )(\beta)}$ results in~\cite{42}
\begin{equation}\label{eq:31} \omega '=\gamma (\omega \mp \Omega),\end{equation}
where the upper (lower) sign refers to positive (negative) helicity radiation. We note
that in the eikonal limit $\Omega /\omega \to 0$ and the instantaneous Doppler result is
recovered. The general problem of electromagnetic waves in a (uniformly) rotating frame
of reference has been treated in~\cite{43}.

It is possible to understand equation~\eqref{eq:31} in terms of  the relative motion of
the observer with respect to the field. In a positive (negative) helicity wave, the
electric and magnetic fields rotate with the wave frequency $\omega\; (-\omega)$ about its
direction of propagation. Thus the rotating observer perceives that the electric and
magnetic fields rotate with frequency $\omega -\Omega\; (-\omega -\Omega)$ about the
direction of wave propagation. Taking due account of time dilation, the observed
frequency of the wave is thus $\gamma (\omega -\Omega)$ in the positive helicity case and
$\gamma (\omega +\Omega )$ in the negative helicity case. These results illustrate the
phenomenon of helicity-rotation coupling for the photon, since~\eqref{eq:31} can be
written as $E'=\gamma (E-\bm{S}\cdot \bm{\Omega})$, where $E=\hbar \omega$,
$\bm{S}=\hbar \hat{\bm{H}}$ and $\hat{\bm{H}}=\pm c\bm{k}/\omega$ is the unit helicity
vector.

It follows from~\eqref{eq:31} that for a slowly moving detector $\gamma \approx 1$ and 
\begin{equation}\label{eq:32} \omega '\approx \omega \mp \Omega ,\end{equation}
which corresponds to the phenomenon of {\em phase wrap-up} in the Global Positioning
System (GPS)~\cite{44}. In fact, equation~\eqref{eq:32} has been verified for $\omega
/(2 \pi)\approx 1$ GHz and $\Omega /(2\pi )\approx 8$ Hz by means of the GPS~\cite{44}.
For $\omega \gg\Omega$, the modified Doppler and aberration formulas due to the
helicity-rotation coupling are~\cite{45}
\begin{align}\label{eq:33} \omega'&=\gamma [(\omega -\hat{\bm{H}}\cdot \bm{\Omega})-\bm{v}\cdot \bm{k}],\\
\label{eq:34}\bm{k}'&=\bm{k}+\frac{1}{v^2}(\gamma
-1)(\bm{v}\cdot\bm{k})\bm{v}-\frac{1}{c^2}\gamma (\omega -\hat{\bm{H}}\cdot
\bm{\Omega})\bm{v},
\end{align}
and similar formulas can be derived for any spinning particle. Circularly polarized
radiation is routinely employed for radio communication with artificial satellites as well
as Doppler tracking of spacecraft. In general, the rotation of the emitter as well as
the receiver should be taken into account. It follows from~\eqref{eq:33} that ignoring
helicity-rotation coupling would lead to a systematic Doppler bias of magnitude
$c\Omega /\omega$. In the case of the Pioneer spacecraft, the anomalous acceleration
resulting from the helicity-rotation coupling has been shown to be negligibly
small~\cite{46}.

A half-wave plate flips the helicity of a photon that passes through it. Imagine a
half-wave plate that rotates uniformly with frequency $\Omega$ and an incident positive
helicity plane wave of frequency $\omega_{\In}$ that propagates along the axis of
rotation. It follows from~\eqref{eq:32} that $\omega '\approx \omega_{\In}-\Omega$. The
spacetime of a uniformly rotating system is stationary; therefore, $\omega '$ remains
fixed inside the plate. The radiation that emerges from the plate has frequency
$\omega_{\out}$ and negative helicity; hence, equation~\eqref{eq:32} implies that $\omega
'\approx \omega_{\out}+\Omega$. Thus the rotating half-wave plate is a frequency shifter: 
$\omega_{\out}-\omega_{\In}\approx -2\Omega$. In general, any rotating spin flipper can
cause an up/down energy shift given by $-2\bm{S}\cdot \bm{\Omega}$ as a consequence of
the spin-rotation coupling. The frequency-shift phenomenon was first discovered in 
microwave experiments~\cite{47} and has subsequently been used in many optical
experiments (see~\cite{45} for a list of references).

Regarding the spin-rotation coupling for fermions, let us note that for experiments in a
laboratory fixed on the Earth, we must add to every Hamiltonian the
spin-rotation-gravity term
\begin{equation}\label{eq:35}\delta H\approx -\bm{S}\cdot \bm{\Omega}+\bm{S}\cdot
\bm{\Omega}_P,\end{equation}
where the second term is due to the gravitomagnetic field of the Earth. That is, the
rotation of the Earth causes a dipolar gravitomagnetic field (due to mass current),
which is locally equivalent to a rotation by the gravitational Larmor theorem. In fact,
$\bm{\Omega}_P$ is the frequency of precession of an ideal fixed test gyro and is given
by
\begin{equation}\label{eq:36}\bm{\Omega}_P\approx \frac{G}{c^2r^5} [3(\bm{J}\cdot
\bm{r})\bm{r}-\bm{J}r^2],\end{equation}
where $J$ is the proper angular momentum of the central source. It follows
from~\eqref{eq:35} that for a spin $\frac{1}{2}$ particle, the difference between the
energy of the particle with spin up and down in the laboratory is characterized by
$\hbar \Omega_\oplus \sim 10^{-19}$eV and $\hbar \Omega_P\sim 10^{-29}$eV, while the
present experimental capabilities are in the $10^{-24}$eV range~\cite{28}. In fact,
indirect observational evidence for the spin-rotation coupling has been
obtained~\cite{48} from the analysis of experiments that have searched for anomalous
spin-gravity interactions~\cite{49}. Further evidence for spin-rotation coupling exists
based on the analysis of muon $g-2$ experiment~\cite{51}.

An experiment to measure directly the spin-rotation coupling for a spin $\frac{1}{2}$
particle was originally proposed in~\cite{32}. This involved a large-scale neutron
interferometry experiment with polarized neutrons on a rotating platform~\cite{52}. A
more recent proposal~\cite{53} employs a rotating neutron spin flipper and hence is much
more manageable as it avoids a large-scale interferometer. The slow neutrons from a
source are longitudinally polarized and the beam is coherently split into two paths that
contain neutron spin flippers, one of which rotates with frequency $\Omega$ about the
direction of motion of the neutrons. In this leg of the interferometer, an energy shift
$\delta H=-2\bm{S}\cdot \bm{\Omega}$ is thus introduced. The two beams are brought back
together and the interference beat frequency $\Omega$ is then measured. It is interesting
to note that a beat frequency in neutron interferometry has already been
measured in another context~\cite{54}; therefore, similar techniques can be used in the
proposed experiment~\cite{53}.

Some general remarks on the calculation of the phase shift are in order here. One starts
from the relation $\hbar\; d\Phi =-Edt+\bm{p}\cdot d\bm{x}$ for the phase
$\Phi (\bm{x},t)$ of the neutron wave in the JWKB approximation. Integrating from the
source $(\bm{x}_S,t_S)$ to the detector $(\bm{x}_D,t_D)$, we find
\begin{equation}\label{eq:37} \hbar \Phi (\bm{x}_D,t_D)=\hbar \Phi
(\bm{x}_S,t_S)-\int^{t_D}_{t_S}Edt+\int^{\bm{x}_D}_{\bm{x}_S}\bm{p}\cdot
d\bm{x}.\end{equation}
Assuming equal amplitudes, the detector output is proportional to
\begin{equation}\label{eq:38} |e^{i\Phi_1}+e^{i\Phi_2}|^2=2(1+\cos \Delta\Phi
),\end{equation}
where $\Phi_1 (\Phi_2)$ refers to the phase accumulated along the first (second) beam
and $\Delta \Phi=\Phi_1-\Phi_2$. It is usually assumed that the two beams are coherently
split at the source; therefore,
\begin{equation}\label{eq:39} \Phi_1 (\bm{x}_S,t_S)=\Phi_2 (\bm{x}_S,t_S).\end{equation}
We thus find
\begin{equation}\label{eq:40} \hbar\; \Delta \Phi =-\int^{t_D}_{t_S}\Delta E\; dt+\oint
\bm{p}\cdot d\bm{x}.\end{equation}
In stationary situations, it is possible to assume that $E_1=E_2=p^2_0/(2m)$, where (for
$i=1,2$)
\begin{equation}\label{eq:41} E_i=\frac{p^2_i}{2m}+\delta H_i.\end{equation}
Thus $\Delta E=0$ and the calculation of the phase shift~\eqref{eq:40} can be simply
performed if the perturbations $\delta H_1$ and $\delta H_2$ are small. it then follows
from~\eqref{eq:41} that if $\delta \bm{p}$ is the perturbation in neutron momentum due
to $\delta H$ such that $\bm{p}-\delta \bm{p}$ is the ``unperturbed" momentum with
magnitude $p_0$, then
\begin{equation}\label{eq:42} \bm{v}\cdot \delta \bm{p}=-\delta H,\end{equation}
where $\bm{v}$ is the neutron velocity. Hence, the \textit{extra} phase shift due to the
perturbation is given by
\begin{equation}\label{eq:43} \Delta \Phi =\frac{1}{\hbar}\oint \delta \bm{p}\cdot
d\bm{x}=\frac{1}{\hbar} \int^D_S(-\delta H_1+\delta H_2)dt.\end{equation}
Consider, as an example, the Sagnac effect in the rotating frame, where
$E=p^2/(2m)+\delta H$ with $\delta H=-\bm{\Omega }\cdot \bm{L}$. Thus
equation~\eqref{eq:43} can be written as $\hbar\; \Delta \Phi =\oint \bm{\Omega}\cdot
(m\bm{r}\times d\bm{r})$, since $\bm{L}=m\bm{r}\times \bm{v}$. In this way, one
immediately recovers equation~\eqref{eq:25}. The approach described here was originally
employed for the calculation of the phase shift due to the spin-rotation coupling in a
uniformly rotating system in~\cite{32}.

In nonstationary situations, such as the proposed experiment using a rotating spin
flipper, $\Delta E\neq 0$ and hence there is a beat phenomenon in addition to a phase
shift. In fact, it follows from the analysis of that experiment~\cite{53} that $\Delta
E=-\hbar \Omega$ for $t>t_{\out}$, when the neutron exits the spin flippers. Hence
$\Delta \Phi$ contains $\Omega (t_D-t_{\out})$ in addition to a phase shift.

It is important to mention briefly the modification of spin-rotation coupling by the
nonlocal theory of accelerated observers (section~\ref{sec:4}). Equation~\eqref{eq:27}
implies that $E'$ can be positive, zero or negative. When $E'=0$, the wave stands
completely still with respect to the static observers in the rotating system. This is
contrary to the basic postulate of the nonlocal theory; therefore, the only modification
in equation~\eqref{eq:27} occurs for the $E'=0$ case. This circumstance is discussed in
detail in~\cite{20}.

\section{Translational Acceleration\label{sec:9}}

Before treating quantum mechanics in translationally accelerated systems, it proves
useful to digress here and discuss the transition from Lorentz invariance to Galilean
invariance in quantum mechanics. What is the transformation rule for a Schr\"odinger
wave function under a Galilean boost $(t=t',\bm{x}=\bm{x}'+\bm{V}t)$? It follows from
Lorentz invariance that for a spinless particle
\begin{equation}\label{eq:44} \phi (x)=\phi '(x'),\end{equation}
where $\phi$ is a scalar wave function that satisfies the Klein-Gordon equation
\begin{equation}\label{eq:45} \left(\Box +\frac{m^2c^2}{\hbar^2}\right)\phi
(x)=0.\end{equation}
To obtain the Schr\"odinger equation from~\eqref{eq:45} in the nonrelativistic limit, we
set
\begin{equation}\label{eq:46} \phi (x)=\varphi
(\bm{x},t)e^{-i\frac{mc^2}{\hbar}t}.\end{equation}
Then,~\eqref{eq:45} reduces to
\begin{equation}\label{eq:47} -\frac{\hbar^2}{2m}\nabla^2\varphi =i\hbar \frac{\partial
\varphi}{\partial t}-\frac{\hbar^2}{2mc^2}\frac{\partial ^2\varphi}{\partial
t^2}.\end{equation}
Neglecting the term proportional to the second temporal derivative of $\varphi$ in the
nonrelativistic limit $(c\to \infty )$, we recover the Schr\"odinger equation for the
wave function $\varphi$.

Under a Lorentz boost,~\eqref{eq:44} and \eqref{eq:46} imply that
\begin{equation}\label{eq:48} \varphi (\bm{x},t)e^{-i\frac{mc^2}{\hbar} t}=\varphi
'(\bm{x}',t')e^{-i\frac{mc^2}{\hbar}t'},\end{equation}
where
\begin{equation}\label{eq:49} t=\gamma \left(t'+\frac{1}{c^2}\bm{V}\cdot
\bm{x}'\right).\end{equation}
It follows from
\begin{equation}\label{eq:50} t-t'=\frac{1}{c^2}\left(\bm{V} \cdot
\bm{x}'+\frac{1}{2}V^2t'\right) +O\left( \frac{1}{c^4}\right)\end{equation}
that in the nonrelativistic limit $(c\to \infty )$,
\begin{equation}\label{eq:51} \varphi (\bm{x},t)=e^{i\frac{m}{\hbar}\left( \bm{V} \cdot
\bm{x}'+\frac{1}{2}V^2t\right)} \varphi '(\bm{x}',t).\end{equation}
This is the standard transformation formula for the Schr\"odinger wave function under a
Galilean boost.

On the other hand, we expect from equations~\eqref{eq:2} and \eqref{eq:44} that in the
absence of spin, the wave function should turn out to be an invariant. Writing
equation~\eqref{eq:48} in the form
\begin{equation}\label{eq:52} \varphi (\bm{x},t)e^{-i\frac{mc^2}{\hbar}t}=[\varphi
'(\bm{x}',t')e^{i\frac{mc^2}{\hbar}(t-t')}]e^{-i\frac{mc^2}{\hbar}t},\end{equation}
we note that the nonrelativisitic wave function may be assumed to be an invariant under
a Galilean transformation
\begin{equation}\label{eq:53} \psi (\bm{x},t)=\psi '(\bm{x}',t),\end{equation}
where
\begin{equation}\label{eq:54} \psi (\bm{x},t)=\varphi (\bm{x},t),\quad \psi
'(\bm{x}',t)=e^{i\frac{m}{\hbar}\left( \bm{V}\cdot \bm{x}'+\frac{1}{2}V^2t \right)}
\varphi '(\bm{x}',t).\end{equation}
That is, in this approach the phase factor in~\eqref{eq:51} that is due to the
relativity of simultaneity belongs to the wave function itself.

The form invariance of the Schr\"odinger equation under Galilean transformations was
used by Bargmann~\cite{55} to show that under the Galilei group, the wave function
transforms as in~\eqref{eq:51}. Bargmann used this result in a thought experiment
involving the behavior of a wave function under the following four operations: a
translation $(\bm{s})$ and then a boost $(\bm{V})$ followed by a translation $(-\bm{s})$
and finally a boost $(-\bm{V})$ to return to the original inertial system. It is
straightforward to see from equation~\eqref{eq:51} that the original wave function
$\varphi (\bm{x},t)$ is related to the final one $\varphi '(\bm{x},t)$ by 
\begin{equation}\label{eq:55} \varphi (\bm{x},t)=e^{-i\frac{m}{\hbar}\bm{s}\cdot \bm{V}}
\varphi '(\bm{x},t).\end{equation}
The phase factor in~\eqref{eq:55} leads to the \textit{mass superselection rule},
namely, one cannot coherently superpose states of particles of different inertial
masses~\cite{55,56}. This rule guarantees strict conservation of mass in nonrelativistic
quantum mechanics. The physical significance of this superselection rule has been
critically discussed by Giulini~\cite{57} and more recently by Greenberger~\cite{58}.
The main point here is that only Lorentz invariance is fundamental, since the
nonrelativistic limit $(c\to \infty)$ is never actually realized.

It should be clear from the preceding discussion that no mass superselection rule is
encountered in the second approach based on the invariance of the wave
function~\eqref{eq:53}. It follows from the hypothesis of locality that the two distinct
methods under discussion here carry over to the quantum mechanics of accelerated
systems~\cite{59}.

Let us therefore consider the transformation to an accelerated system
\begin{equation}\label{eq:56} \bm{x}=\bm{x}'+\int^t_0\bm{V}(t')dt',\end{equation}
where $\bm{a}=d\bm{V}/dt$ is the translational acceleration vector. Starting from the
Schr\"odinger equation $H\psi =i\hbar \partial \psi /\partial t$ and assuming the
invariance of the wave function, $\psi (\bm{x},t)=\psi '(\bm{x}',t)$, as in the second
approach, we find that $\psi '(\bm{x}',t)=U\psi (\bm{x}',t)$, where
\begin{equation}\label{eq:57} U=e^{\frac{i}{\hbar}\int^t_0\bm{V}(t')\cdot
\bm{p}\; dt'}.\end{equation}
If follows that $\psi '$ satisfies the Schr\"odinger equation $H'\psi'=i\hbar \partial
\psi '/\partial t$ with the Hamiltonian
\begin{equation}\label{eq:58} H'=UHU^{-1}-\bm{V}(t)\cdot \bm{p},\end{equation}
where $\bm{p}$ is the invariant canonical momentum. Writing $H=p^2/(2m)+W$, where $W$ is
the invariant potential energy, we find
\begin{equation}\label{eq:59} \left[
\frac{1}{2m}(\bm{p}-m\bm{V})^2-\frac{1}{2}mV^2+W\right] \psi '=i\hbar \frac{\partial
\psi '}{\partial t}.\end{equation}
Let
\begin{equation}\label{eq:60} \psi '(\bm{x}',t)=e^{i\frac{m}{\hbar}\left[\bm{V}\cdot
\bm{x}' +\frac{1}{2}\int^t_0V^2(t')dt'\right]}\varphi '(\bm{x}',t),\end{equation}
then $\varphi '(\bm{x}',t)$ satisfies the Schr\"odinger equation
\begin{equation}\label{eq:61} \left( -\frac{\hbar^2}{2m}{\nabla '}^2 +m\bm{a}\cdot
\bm{x}'+W\right) \varphi '=i\hbar \frac{\partial \varphi '}{\partial t},\end{equation}
where $\bm{\nabla }'=\bm{\nabla}$ follows from~\eqref{eq:56}. It is important to
recognize that $\varphi '(\bm{x}',t)$ is the wave function from the standpoint of the
accelerated system according to the first (Bargmann) approach. Here the acceleration
potential $m\bm{a}\cdot \bm{x}'$, where $-\bm{\nabla}'(m\bm{a}\cdot \bm{x}')=-m\bm{a}$ is
the inertial force acting on the particle, corresponds to the inertial term that appears
in~\eqref{eq:19}. The existence of this inertial potential has been verified
experimentally by Bonse and Wroblewski~\cite{60} using neutron interferometry. In
connection with the problem of the wave function in the accelerated system---i.e.
whether it is $\varphi '$ or $\psi'$---a detailed examination of the experimental
arrangement in~\cite{60} reveals that this experiment cannot distinguish between the two
methods that differ by the phase factor given in equation~\eqref{eq:60}. Specifically,
the interferometer in~\cite{60} oscillated in the horizontal plane and the intensity of
the outgoing beam was measured at the inversion points of the oscillation at which the
magnitude of acceleration was maximum but $\bm{V}=0$; therefore, the phase factor in
question was essentially unity. To conclude our discussion, it is interesting to
elucidate further the physical origin of this phase factor using classical
mechanics~\cite{32}.

Under the transformation~\eqref{eq:56}, $\bm{v}=\bm{v}'+\bm{V}(t)$ and the Lagrangian of
a classical particle $L=\frac{1}{2}mv^2-W$, with $L(\bm{x},\bm{v})=L'(\bm{x}',\bm{v}')$,
becomes $L'=\frac{1}{2}m(\bm{v}'+\bm{V})^2-W$ in the accelerated system. In classical
mechanics, there are two natural and equivalent ways to deal with this Lagrangian. The
first method consists of writing~\cite{31}
\begin{equation}\label{eq:62} L'=\frac{1}{2}m{v'}^2-m\bm{a}\cdot
\bm{x}'-W+\frac{dF}{dt},\end{equation}
where $F$ is given, up to a constant, by
\begin{equation}\label{eq:63} F=m\bm{V}(t)\cdot
\bm{x}'+\frac{1}{2}m\int^t_0V^2(t')dt'.\end{equation}
The total temporal derivative in~\eqref{eq:62} does not affect the classical dynamics in
accordance with the action principle and hence we confine our attention to
$L'_1=\frac{1}{2}m{v'}^2-m\bm{a}\cdot \bm{x}'-W$. The momentum in this case is
$\bm{p}'=m\bm{v}'$ and the Hamiltonian is thus given by
\begin{equation}\label{eq:64} H'_1=\frac{{p'}^2}{2m}+m\bm{a}\cdot
\bm{x}'+W,\end{equation}
which corresponds to the Hamiltonian in the Schr\"odinger equation~\eqref{eq:61}. The
second method deals with $L'$ without subtracting out $dF/dt$. In this case, the
momentum is the invariant canonical momentum $\bm{p}=m(\bm{v}'+\bm{V})$ and the
Hamiltonian is
\begin{equation}\label{eq:65} H'=\frac{p^2}{2m}-\bm{p}\cdot \bm{V}+W,\end{equation}
which corresponds to equation~\eqref{eq:58} and the Hamiltonian in the Schr\"odinger
equation~\eqref{eq:59}.

In classical mechanics, the two methods represent the same dynamics. Quantum
mechanically, however, there is a phase difference, which can be easily seen from the
path integral approach. That is,
\begin{equation}\label{eq:66} \psi '(\bm{x}',t)=\Sigma
e^{\frac{i}{\hbar}\mathcal{S}'},\end{equation}
where $\mathcal{S}'$ is the classical action,
\begin{equation}\label{eq:67} \mathcal{S}'=\int L'(\bm{x}',\bm{v}')dt.\end{equation}
It follows from~\eqref{eq:62} that
\begin{equation}\label{eq:68} \mathcal{S}'=\mathcal{S}'_1+F,\end{equation}
where $\mathcal{S}'_1$ is the action corresponding to $L'_1$. Using~\eqref{eq:68} and
the fact that
\begin{equation}\label{eq:69} \varphi '(\bm{x}',t)=\Sigma
e^{\frac{i}{\hbar}\mathcal{S}'_1},\end{equation}
we find
\begin{equation}\label{eq:70} \psi '(\bm{x}',t)=e^{\frac{i}{\hbar}F}\varphi
'(\bm{x}',t),\end{equation}
in agreement with equation~\eqref{eq:60}.

It would be interesting to devise an experiment of the Bonse-Wroblewski~\cite{60} type
that could distinguish between the two methods and hence remove the phase ambiguity in
the treatment of translationally accelerated systems.

\section{Discussion\label{sec:10}}

The main observational consequences of Dirac's equation in noninertial frames of
reference are related to the Sagnac effect, the spin-rotation coupling and the
Bonse-Wroblewski effect. These inertial effects can be further elucidated by
interferometry experiments involving matter waves. In particular, a neutron
interferometry experiment has been proposed for the direct measurement of inertial
effect of intrinsic spin. Moreover, neutron interferometry experiments involving
translationally accelerated interferometers may help resolve the phase ambiguity in the
description of the wave function from the standpoint of a translationally accelerated
system.

\end{document}